\documentclass[a4paper, 11pt]{article}

\usepackage{graphicx}
\usepackage{listings}
\usepackage{amsmath, amssymb}

\begin{document}
\title{Proof of Authenticity of Logistics Information with Passive RFID Tags
and Blockchain}
\author{Hiroshi Watanabe\footnote{Department of Electrical and Computer
Engineering, National Chiao Tung University},
Kenji Saito\footnote{Graduate School of Business and Finance,
Waseda University},
Satoshi Miyazaki\footnote{LAPIS Technology Co., Ltd.},
Toshiharu Okada\footnotemark[3],\\
Hiroyuki Fukuyama\footnotemark[3],
Tsuneo Kato\footnote{Beyond Blockchain Inc.},
Katsuo Taniguchi\footnotemark[4]
}
\date{}

\maketitle

\begin{abstract}
In tracing the (robotically automated) logistics of large quantities of goods,
inexpensive passive RFID tags are preferred for cost reasons.
Accordingly, security between such tags and readers have primarily been studied
among many issues of RFID.
However, the authenticity of data cannot be guaranteed if logistics services
can give false information.
Although the use of blockchain is often discussed, it is simply a recording
system, so there is a risk that false records may be written to it.

As a solution, we propose a design in which a digitally signing,
location-constrained and tamper-evident reader atomically writes an evidence to
blockchain along with its reading and writing a tag.

By semi-formal modeling, we confirmed that the confidentiality and integrity of
the information can be maintained throughout the system, and digitally signed
data can be verified later despite possible compromise of private keys or
signature algorithms, or expiration of public key certificates.
We also introduce a prototype design to show that our proposal is viable.

This makes it possible to trace authentic logistics information using
inexpensive passive RFID tags.
Furthermore, by abstracting the reader/writer as a sensor/actuator, this
model can be extended to IoT in general.

\end{abstract}

\section{Introduction}
\subsection{Motivation}
Today, the logistics of physically moving resources and materials has become
increasingly important, as behaviors of people change, and society is
restructured through the penetration of digital communication technologies,
the spread of infectious diseases and other factors, and as remote presence
and remote work become more common.

In tracing the (robotically automated) logistics of large quantities of goods,
inexpensive passive RFID (Radio Frequency IDentification) tags are easy to
use from a cost perspective.
Accordingly, security between such tags and readers have primarily been
studied in the literature among many issues surrounding RFID in the past,
as we will describe in section~\ref{subsec-rfid}.
However, the authenticity of the identified information cannot be guaranteed
by measures between tags and readers alone if the communication between
clients and services is considered, because services may give false
information, deliberately or by accident.
Although the use of blockchain is often discussed, it is simply a recording
system, so there is a risk that false records may be written to it at the
beginning, in which case the tamper-evidentness of blockchain becomes
meaningless.

We feel a strong need to overcome these challenges, and ensure the
authenticity of information obtained through RFID in a logistics network.

\subsection{Contributions}
Contributions of this work are the following:
\begin{enumerate}
\item We proposed a design of an information infrastructure in which the
information on tags read and/or written at controlled locations by controlled
RFID readers are correctly shared in the logistics network through use of
blockchain.
\item We made a series of propositions that need to be true in order for such
a design to work as intended.
We made a semi-formal model of our proposal, and verified that all the
propositions are true under the model.
\item We introduced a prototype design, and evaluated feasibility of the
design, especially with respect to usage of blockchain in reality.
\end{enumerate}

\subsection{Organization of This Paper}
The rest of this paper is organized as follows.
In section~\ref{sec-background}, we describe the technology underlying our
proposal: RFID and its security, tamper-evidentness, cryptography and
blockchain.
Section~\ref{sec-problem} defines the problem and propositions that need to
be true under any solutions.
Section~\ref{sec-design} introduces our proposed design and its semi-formal
model, with which we verify the correctness of the design.
Section~\ref{sec-feasibility} evaluates the feasibility of our proposal through
a prototype design.
Section~\ref{sec-discussion} discusses applications and generalization of
our proposal.
Section~\ref{sec-related-work} compares our proposal with other related
work.
Finally, section~\ref{sec-conclusions} gives conclusive remarks.

\section{Background}\label{sec-background}

\subsection{RFID and Its Security}\label{subsec-rfid}
RFID\cite{Razaq4588226} is usually used in a form of exchanging data stored
in IC tags by wireless communication using radio waves.
An IC tag is composed of an IC chip with small-bit capacity semiconductor
memory and antenna.
The stored data is editable, which is different from printable codes such as
barcode or QR (Quick Response) code.

The frequency bands are categorized as the low frequency (LF), the high
frequency (HF), the ultra-high frequency (UHF) and microwaves. 
This work assumes the UHF band, which employs about 900 MHz.
In the case, the communication is done by using electronic field, and the
communication distance is several meters.
It is suitable to collect multiple set of information from our belongings
or numerous items in a cardboard box simultaneously.

The other categorization is with or without battery.
An {\em active} RFID tag has battery inside, and can thus emit radio wave by
its own.
The communication distance is comparatively long (1 to 100 meters), and it is
suitable to sensors networks, but its service life is limited by battery
lifetime.
On the other hand, a {\em passive} RFID tag can function acquiring power from
the reader's radio wave.
It is generally inexpensive, and can work semi-permanently,
i.e., free from battery lifetime, although the communication distance is
comparatively short.
This work assumes passive RFID tags, because they can be deployed
inexpensively throughout a large-scale logistics network.

\begin{figure}[h]
\begin{center}
\includegraphics[scale=0.6]{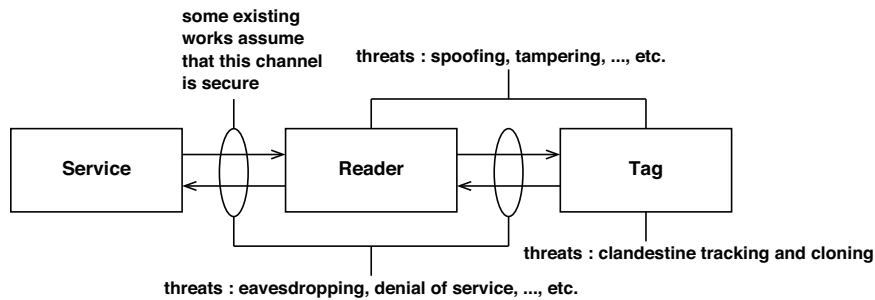}\\
\caption{RFID Security and Privacy Threats}\label{fig-rfid-threats}
\end{center}
\end{figure}

Privacy and counterfeiting are two major concerns in the security issues of
RFID, including clandestine tracking, eavesdropping, spoofing and cloning,
extensibly studied in \cite{Juels1589116} and \cite{Koscher1653668}.
Fig.~\ref{fig-rfid-threats} shows possible threats.
To tackle these issues, mutual authentication protocol among the service,
reader, and tag has been studied in \cite{Aghili2019621} and
\cite{Safkhani2020107558}, for example.
Their correctness has been analyzed in formal logics such as
BAN\cite{Burrows1990:BAN} and/or GNY\cite{Gong63854}.

\subsection{Tamper Evidence}
To deal with the problem of counterfeiting or cloning, protection against
tampering is necessary.
Such protection by hardware is assumed for both the tag and the reader in this
work.
In particular, we assume {\em tamper evidentness}, which means that if the
hardware is attacked, at least the unerasable evidence of the attack remains
in the hardware.

Such tamper-evident technology already exists in the context of RFID.
For example, \cite{Sadeghi2020:PUF} proposes use of PUF (Physically Unclonable
Function) that provides fingerprints of chips based on their physical
properties, in order to realize tamper-evident storage of cryptographic secrets
for tags.
\cite{Lindsay2006:Evident} proposes to physically deactivate the transmit
capability of a tag upon tampering the package.
There are also studies to make microprocessors\cite{Waksman5504715} or the
whole architecture for processing\cite{Suh2003:AEGIS} tamper-evident.

\subsection{Cryptographic Hash Function}
To provide an evidence to prove or disprove the existence of a readout from
a tag, a {\em cryptographic hash function} is used in this work.
Such a function can take data of any size, of any type as input, and returns a
value of a fixed bit length, such as 256bits, as determined by each function.
The output value is often called a {\em digest}.
The input data ({\em preimage}) cannot be deduced from the digest, so if the
preimage is different by even one bit, the digest to be obtained will be
completely different.
Because the preimages in an infinite space are mapped to a finite albeit vast
space with a fixed bit length, it is theoretically possible to obtain equal
digests from different preimages.
This is called a {\em collision}.
Probability of an occurrence of a collision is virtually negligible.

SHA-3 (Secure Hash Algorithm-3) is an example of cryptographic hash functions
in actual use, which is based on Keccak\cite{Bertoni2008:Keccak} algorithm
used in some blockchain.

\begin{figure}[h]
\begin{center}
\includegraphics[scale=0.38]{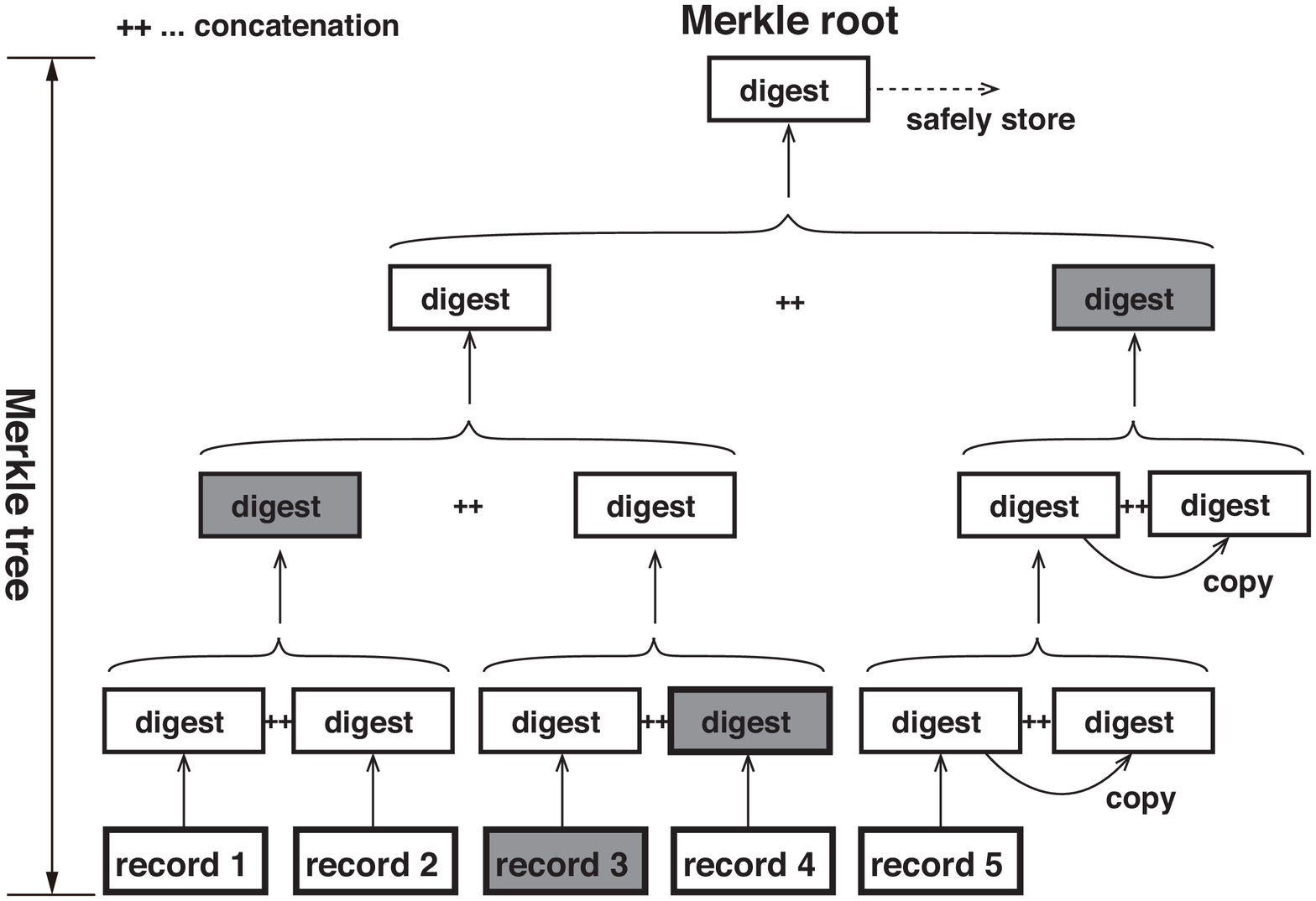}\\
{\footnotesize
\begin{itemize}
\item[*] In order to confirm the existence of record 3, see if the same
{\em Merkle root} as safely stored can be calculated from the provided partial
tree ({\em Merkle proof}) shown in gray.
\end{itemize}
}
\caption{Merkle tree and Merkle proof}\label{fig-merkle}
\end{center}
\end{figure}

Fig.~\ref{fig-merkle} shows a hash tree data structure and algorithm called
{\em Merkle tree}\cite{Merkle1988:Tree}, with which existence of any one
of numerous records can be tested through one single
digest (root of the tree) safely stored and a partial tree.

\subsection{Digital Signature}
Digital signature is an application of public key cryptography,
and is a technique for proving the authenticity of digital information.
In our proposal, a readout from a tag is digitally signed by the reader.

Given a key pair $\langle k, k^{-1} \rangle$ where $k^{-1}$ is the
{\em private key} that must be kept secret by the user themselves (in the case
of our proposal, the reader itself), and $k$ is the corresponding
{\em public key} that can be made public, we define signing and verifying
operations over some {\em plain text} (data) as follows:
\begin{enumerate}
\item {\em Signing} takes two arguments, plain text and private key $k^{-1}$,
and produces a {\em signature}, a binary sequence.

\item {\em Verifying} takes three arguments, the plain text, the signature and
public key $k$, and produces either of the following:
\begin{itemize}
\item {\em OK} if it is inferred that $k^{-1}$ is used upon signing
(but $k^{-1}$ itself is never revealed), and that the plain text has not
changed since the signature is made.
\item {\em NG (No Good)} otherwise.
\end{itemize}
\end{enumerate}

Some examples of digital signature algorithms in actual use include
RSA\cite{RSA1978:Signature},
ECDSA\cite{Johnson2001:ECDSA}, and
Ed25519\cite{Bernstein2012:Ed25519}.

For a digital signature to be used as a proof of something as intended,
we must first prove the authenticity of the public key $k$, or otherwise the
signature may have been spoofed.
X.509\cite{Cooper2008:X509} defines public key certificates for that purpose,
requiring a structure of CAs (Certificate Authorities) who digitally sign such
certificates transitively, and is a standard used on the Internet.
Expiration of a public key certificate is often a risk that makes digital
signatures unverifiable.
Other risks are compromising the private key and compromise of the signature
algorithm itself, in either case digital signatures can be forged.

These risks make it rather difficult to maintain the proving power of digital
signatures over the long term.
\cite{Toyoshima2005:NICT} summarizes this problem as two sub-problems that are
two sides of the same coin:
\begin{description}
\item[Elapsed-time proof problem] asks whether it can be proven
that the signature is not a fake and that the presented digital signature was
created before a specific point in the past
(e.g., when the private key was compromised or the compromise of the signature
algorithm became apparent).

\item[Alibi proof problem] asks whether it can be proven
that the signature is a fake and that the presented digital signature did not
exist at the particular time in the past.
\end{description}
Both sub-problems assume that the private key or the signature algorithm can
be compromised, or the public key certificate can be revoked or expired.
Because in a large-scale network some certificates are expiring at any moment,
and we may need to prove authenticity of data in the past,
these problems are real for this work.

\subsection{Blockchain}\label{subsec-blockchain}
We use blockchain to solve elapsed-time and alibi proof problems of digital
signatures to provide firm evidences of RFID readouts.
Blockchain was invented as part of Bitcoin\cite{Nakamoto2008:Bitcoin}, to
enable users to transfer their funds without interference from anyone,
including governments or banking institutions.
Blockchain was designed as the foundation for actualizing this goal, and is
expected to realize a state machine that satisfies the following properties
(BP: Blockchain Properties), as described in \cite{Saito2020:Blockchain}:
\begin{description}
\item[BP-1:] Only an authorized user can cause a state transition that is
allowed in the state machine.
\item[BP-2:] Such a state transition always occurs if the authorized user
wants it to happen.
\item[BP-3:] Once a state transition occurs, it is virtually irreversible.
\end{description}

\begin{figure}[h]
\begin{center}
\includegraphics[scale=0.4]{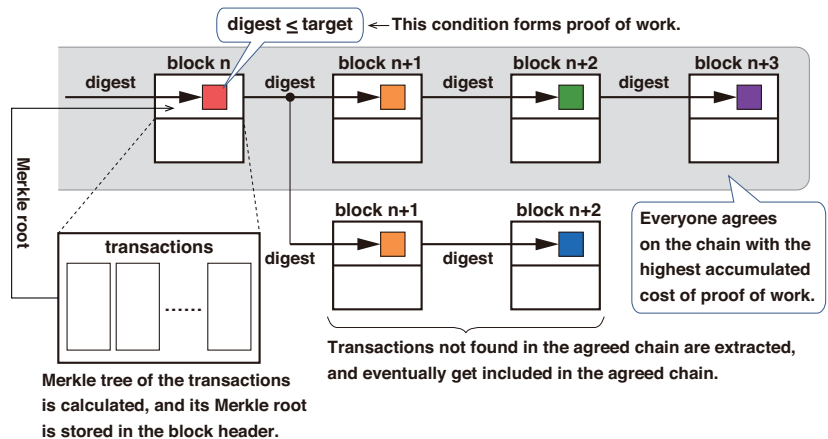}\\
\caption{Blockchain Based on Proof of Work}\label{fig-blockchain}
\end{center}
\end{figure}

Fig.~\ref{fig-blockchain} illustrates the structure of blockchain based on
{\em proof of work}, which has been employed in Bitcoin and many other
cryptocurrencies.
Each block of transactions contains the cryptographic digest of the previous
block, except the very first block sometimes called the genesis block.
Such a digest must meet a certain criterion; it needs to be less than or equal
to the pre-adjusted and agreed target stored in or calculated from the block.
Since the digest is calculated by a one-way function whose outputs are evenly
distributed, no one can intentionally configure a block to satisfy the
criterion.
Instead, they need to partake repetitive trials to change the values of nonce
in the block they are creating until they get a right digest, which will be
stored in the next block as the proof of performed work (thus we call this
structure {\em hash chain with proof of work}).

The necessity of repetitive trials functions as a protection against
falsification.
A transaction itself cannot be falsified unless digital signatures are
compromised.
But it is conceivable to remove some transactions from a past block, or to add
fabricated transactions that did not really exist.
If one tries so, the digest of the block is changed, and is typically greater
than the target.
Then they would have to retry the proof of work for the block.
This changes the digest stored in the next block, which in turn means that the
digest of the next block is also changed, and is typically greater than the
target, and so on.
In short, ones with a malicious intention would have to redo the proof of work
from where they want to change, and outdo the ongoing process of adding blocks
eventually to make the change valid, which has generally been considered
highly difficult.

There is a possibility of multiple participants each proposing a new block at
roughly the same time, which may be accepted by different sets of participants.
Then the hash chain may have multiple ends that are extended independently
from one another, resulting in a fork of the blockchain with multiple
(and possibly, contradicting) histories of blocks.
If this happens, the branch that is the most difficult to produce is chosen by
all participants ({\em Nakamoto consensus}).
This reflects the total cost cast in the creation of the hash chain branch;
because of proof of work, any chain branch requires the same cost paid for
its creation when it is tried to be falsified.

Ethereum\cite{Buterin2013:Ethereum} is an application platform based on
blockchain, where applications are abstracted as so-called
{\em smart contracts} communicating one another, whose code, execution logs,
and the resulted states are all recorded onto blockchain, and are verifiable
through Merkle proof.
Deploying and running smart contracts cost {\em gas}, the representation of
virtual CPU cycles paid by ETH, the native currency of Ethereum.
We are using Ethereum as the blockchain platform in our prototype design
as to be described in section~\ref{sec-feasibility}.

Even if blockchain itself is tamper-evident, and it is difficult to tamper
with its records without introducing some inconsistency, it cannot be detected
if the records fed in from the external world are themselves false.
This is known as {\em oracle problem}\cite{Adler8726819}, and is one of the
issues we are trying to solve in this work.

\section{Problem}\label{sec-problem}
\begin{figure}[h]
\begin{center}
\includegraphics[scale=0.6]{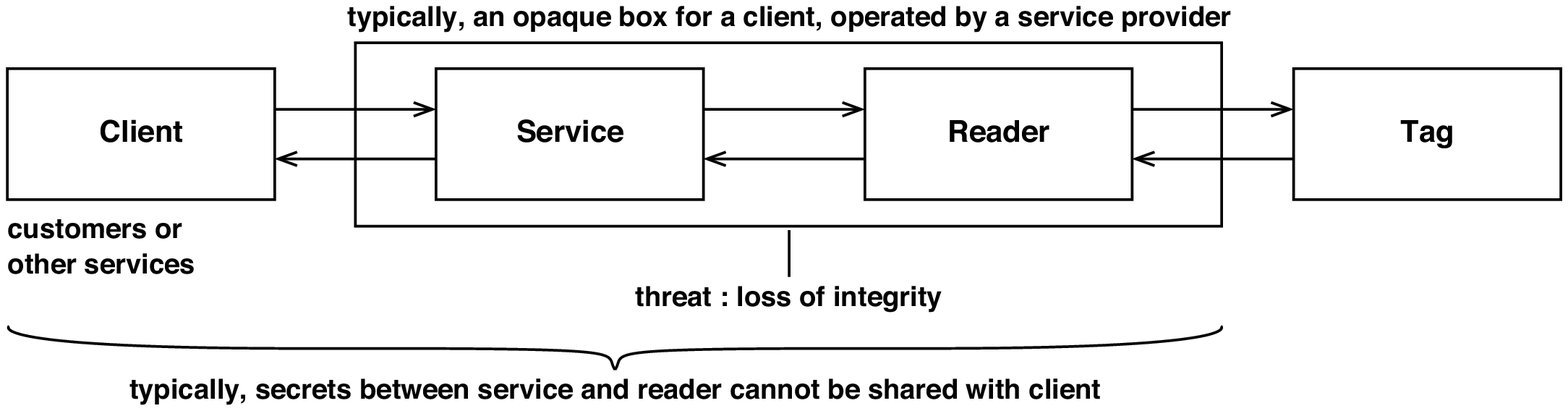}\\
\caption{Problem}\label{fig-problem}
\end{center}
\end{figure}

We would like to design an information infrastructure to ensure that
information written to or read from tags at specified locations by the
controlled RFID readers is correctly shared in the logistics network.
In particular, as illustrated in Fig.~\ref{fig-problem}, we have to solve the
problem of whether the data in the tag read and written by the reader is
treated with integrity, and communicated with the client as it is, even
through the service provider that can be seen as an opaque box.
Security between the reader and the tag is out of scope of this work, although
unclonability of tags is assumed.

To solve the problem, we will introduce an additional entity to provide
digital evidences, and thus all propositions below need to be true under the
proposed solution:
\begin{enumerate}
\item {\em Intra-service confidentiality} : No secret between the service and
the reader is shared with the client, yet the client can verify the
authenticity of data from the reader.
\item {\em Service integrity} : If the service lies about the content or
existence of data, it is detected by the client.
\item {\em Service confidentiality} : No secret between the service and the
client is shared with the additional entity that provides digital evidences.
\item {\em Evidence service integrity} : If the additional entity lies about
the content or existence of a digital evidence, it is detected by the client.
\item {\em Elapsed-time proof} : Past data remains authentic even after
related private keys or the digital signature algorithm are compromised, or
public key certificates are revoked or expired.
Furthermore, any attempt to delete past authentic data will fail.
\item {\em Alibi proof} : Non-existent data in the past cannot be fabricated
as if it existed, even after related private keys or the digital signature
algorithm are compromised.
\end{enumerate}

\section{Design}\label{sec-design}
\subsection{Outline}
\begin{figure}[h]
\begin{center}
\includegraphics[scale=0.6]{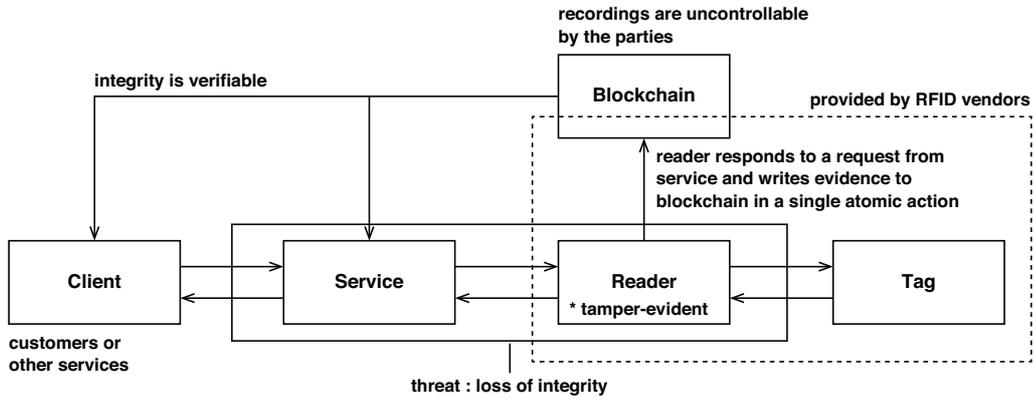}\\
\caption{Overview of the Design}\label{fig-design}
\end{center}
\end{figure}

Our design is outlined as Fig.~\ref{fig-design}.

In our proposal, an RFID reader, identified by digital signatures, functionally
constrained to work only at a specific physical location through a positioning
system, and is tamper-evident, writes the digital evidence to blockchain as it
reads from and writes to the RFID tag in an atomic (non-separable) operation.
To be tamper-evident, a reader is assumed to be equipped with an unclonable
chip, and needs to be occasionally replaced by its vendor as the public key
certificate expires.

\begin{figure}[h]
\begin{center}
\includegraphics[scale=0.6]{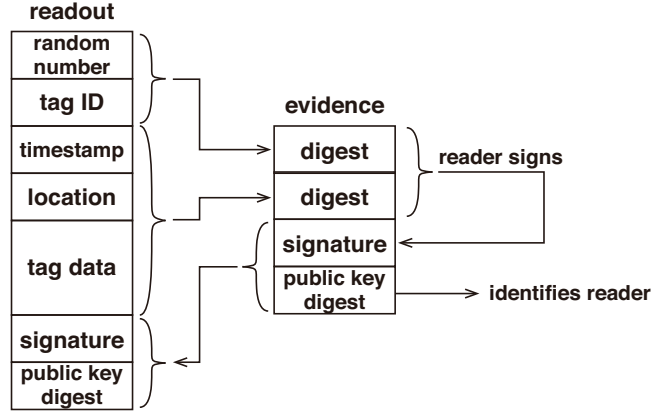}\\
\caption{Example Format of Reader Readout and Evidence}\label{fig-evidence}
\end{center}
\end{figure}

RFID data readout and its evidence are formulated as illustrated in
Fig.~\ref{fig-evidence}.
A public key digest is used as the identifier of a reader.
A $\langle$random number, tag ID$\rangle$ pair is used as a key for searching
readouts or evidences.
The random number is used for hiding the tag ID from blockchain operators,
and is shared among clients, the service and readers (presumably generated by
the service).
Timestamp and location are assumed to be obtained by the reader's internal
positioning system.

\subsection{Semi-formal Model}

We present a more formal model to describe the details.

\subsubsection{Logistics Network}
A logistics network is a tuple
$\langle E, V, R, C, L, B, T \rangle$
where $E$ is the set of participating entities (services and customers),
$V$ is the set of RFID reader vendors,
$R$ is the set of RFID readers,
$C$ is the set of RFID tag chips and packages represented by the tags,
$L$ is the set of locations, and
$B$ is the set of participating nodes in blockchain.
$T$ is the set of time, values of whose elements monotonously increase along
with the direction of time, so that for $t_1, t_2 \in T$,
if $t_1$ precedes $t_2$ then $t_1 < t_2$.
Also, $t_1 \approx t_2$ represents a timewise proximity between the two points
of time, and $t_1 \nsim t_2$ represents a timewise remoteness.
$t_1 \lessapprox t_2$ and $t_1 \lnsim t_2$ represent that the two points of
time $t_1 < t_2$ are close and apart, respectively\footnote{The specifics
for closeness and apartness of time depend on the blockchain system,
but in the case of Ethereum, where the average block interval is 15 seconds,
we can say that times up to a few tens of seconds are close and minutes or
more are apart.}.

\subsubsection{Communication}
Let $A$ be the set of {\em terms}.
For all $a_2 \in A$, if $a_1$ is part of $a_2$, then $a_1 \in A$, and the
relation is denoted as $a_1 \sqsubset a_2$, i.e., $a_1$ is a {\em subterm} of
$a_2$.
Also, $a_1 \parallel\sqsubset a_2$ denotes $a_1$ is one of disjoint subterms of
$a_2$.
For any $A' \subset A$ such that $|A'| = n$ where $n > 0$, a concatenation of
the elements of $A'$ is denoted as $\langle a_1, a_2, \dots, a_n \rangle$,
which is also in $A$.
Terms can be sent and received via a communication channel.
$L \subset A$ and $T \subset A$.

Between two communication-capable elements $x$ and $y$ of sets in the
logistics network, there can be communication relations such that
$x \xrightarrow[t]{a} y$ and $y \xleftarrow[t]{a} x$, which both mean
$x$ sends $a$ to $y$ at time $t$, where $a \in A$ and $t \in T$.

$x \xrightarrow[t]{a} y$ implies that $x$ {\it knows} $a$ at time $t$, and
$y$ {\it knows} $a$ at all $t'$ such that $t' \geq t$ unless
$y$ {\it forgets} $a$ at some point.
If $x$ {\it knows} $a$ at time $t$, then $x$ {\it knows} all $a'$ at $t$ such
that $a' \sqsubset a$.
If $x$ {\it forgets} $a$ at time $t$ such that $a \sqsubset a'$, then $x$
{\it forgets} $a'$ also at time $t$.

\subsubsection{Cryptography}
$H$ is a cryptographic hash function.
For all $x$, $H(x) \in A$.

$H(\langle x_1, x_2, \dots, x_n \rangle)$, where $\langle \dots \rangle$
represents a concatenation, is denoted as $H(x_1, x_2, \dots, x_n)$ for
simplicity.

$\langle k, k^{-1} \rangle$ denotes a key pair.
A digital signature by using a private key $k^{-1}$ over term $a \in A$ is
denoted as $sig(a, k^{-1})$.
To verify the signature, $a$ and the corresponding public key $k$ are required.
$sig(a, k^{-1}) \in A$.

\subsubsection{Tags}
A tag is identified by an unclonable function $U_{tag} : C \rightarrow EPC$
where $EPC$ is the set of electronic product code, and $EPC \subset A$.
Because of unclonability of the function, it is extremely difficult to have
$c'$ such that $c' \neq c$ and $U_{tag}(c') = U_{tag}(c)$.
$U_{tag}(c)$ is also denoted as $c.id$,$\:$ i.e., the identifier of $c$.

There is a memory readout function $MRO : C \times T \rightarrow D$,
where $D$ is the set of possible data, and $D \subset A$.
$MRO(c,t)$ is also denoted as $c.data(t)$, i.e., data read from the memory of
tag $c$ at time $t$.
If $c.data(t) = d$, then $c.data(t') = d$ for all $t'$ such that $t < t'$,
unless another data is written at some point.

There also is a physical location function
$LF_{physi} : C \times T \rightarrow L$.
$LF_{physi}(c,t)$ is also denoted as $c.loc(t)$, i.e.,
the physical location of tag $c$ at time $t$.

\subsubsection{Readers}
A reader is associated with an unclonable function
$U_{reader} : R \rightarrow K$ where $K$ is the set of cryptographic keys.
Because of unclonability of the function, it is extremely difficult to have
$r'$ such that $r' \neq r$ and $U_{reader}(r') = U_{reader}(r)$.
$U_{reader}(r)$ is the private key of $r$ that is also denoted as $k_r^{-1}$,
from which the corresponding public key $k_r$ is calculated.
$r$ is identified by $H(k_r)$.

Let $E_{serv}$ be the set of logistical services such that
$E_{serv} \subset E$.
A function $OwnerOf: R \rightarrow E_{serv}$ maps readers to services to which
they belong.

For simplicity, every $r$ is situated at a logistical location $l$
(e.g., a relay point in the logistics network) such that $l \in L_{logi}$ and
$L_{logi} \subset L$.
There is a logistical location function $LF_{logi} : R \rightarrow L_{logi}$.
$LF_{logi}(r)$ is also denoted as $r.loc()$, i.e.,
the logistical location of reader $r$.
If reader $r$ reads or writes to tag $c$, that is, 
$r \xleftarrow[t]{a} c$ or $r \xrightarrow[t]{a} c$ for some $a \in A$ and
$t \in T$, then $r.loc() \approx c.loc(t)$, representing a locationwise
proximity.

A read operation by $r$ against tag $c$ at time $t$ is expressed as
$r \xleftarrow[t]{\langle c.id,c.data(t) \rangle} c$, and
a write operation is expressed as
$r \xrightarrow[t]{d} c \implies c.data(t) = d$.
In either case, $r$ {\it knows} $\langle c.id, c.data(t) \rangle$ at time $t$,
and $r$ {\it forgets} $\langle c.id, c.data(t) \rangle$ at some time $t'$ such
that $t \lnsim t'$.

Also upon such a read or write operation, $r$ reports {\em readout} and
{\em evidence} as follows:
Let $n \in A$ be a predefined random number shared with service
$s = OwnerOf(r)$ and its clients,
$h_1 = H(n, c.id)$, and $h_2 = H(t, r.loc(), c.data(t))$.
Then,
\begin{description}
\item [readout]
$ro = \langle n, c.id, t, r.loc(), c.data(t),
sig(\langle{}h_1, h_2\rangle, k_r^{-1}), H(k_r) \rangle$

\item [evidence]
$ev = \langle h_1, h_2, sig(\langle h_1, h_2 \rangle, k^{-1}_r),
H(k_r) \rangle$

\end{description}

\paragraph{Atomicity of Readout and Evidence}
On the above $ro$ and $ev$,
$r\xrightarrow[t_{ro}]{ro}s$ if and only if $r\xrightarrow[t_{ev}]{ev}b$
where $b \in B$, $t < t_{ro}$, $t < t_{ev}$,
and $t_{ro} \approx t_{ev}$.

Let $RO^s_t$ be the set of all readout $ro$ that $s$ {\it knows} at time $t$,
and $EV^b_t$ be the set of all evidence $ev$ that $b$ {\it knows} at $t$.
Furthermore, let $EV$ be the set of all evidences, and $EV_t$ be the set of
all evidences that all correct (neither faulty nor malicious) $b \in B$
collectively {\it knows} at $t$.

\paragraph{Tamper Evidentness}
Any attempt to tamper with $r$ at some time $t$ would result in change of $r$'s
physical characteristics, such that $k_r^{-1}$ at $t$ differs from that at
any $t'$ such that $t' < t$.

\subsubsection{Reader Vendors}
Let $R_v \subset R$ be the set of readers vended by vendor $v \in V$.
$v$ issues a certificate for all $r \in R_v$ to prove authenticity of $r$'s
public key $k_r$.

\begin{description}
\item[certificate]
$cert = \langle k_r, t_{ini}, t_{exp},
sig(\langle k_r, t_{ini}, t_{exp} \rangle, k_v^{-1}),
H(k_v)\rangle$, where $cert$ is only valid since time $t_{ini}$ and until
time $t_{exp}$.
\end{description}

It is assumed that $k_v$ and its authenticity to verify the signature is
provided and certified through some external PKI (Public Key
Infrastructure)\footnote{Use of an external PKI makes this model somewhat weak,
as their certificates are not guaranteed for elapsed-time or alibi proofs.}.
For simplicity, we abstract certificate revocation as an early expiration time
in this model.

Upon creation of $cert$ at time $t$ (typically before shipment of $r$),
which is at or before $t_{ini}$,
$v \xrightarrow[t]{cert} b$ where $b \in B$.
$cert$ is subject to recording in blockchain.
$CERT^b_t$ is the set of all certificate $cert$ that $b$ {\it knows} at $t$,
$CERT$ is the set of all certificates, and $CERT_t$ is the
set of all certificates that all correct $b \in B$ collectively {\it knows} at
$t$.

\subsubsection{Reader/Vendor Validation}
When $x \in E \cup B$ {\it knows} term $a$ at time $t$ such that
$a = \langle a_1, sig(a_2, k^{-1}), H(k)\rangle$, where either $a_1 = a_2$
or $a_2 = \langle H(a') \:|\: a' \parallel\sqsubset a_1\rangle$
(this applies to a readout, evidence and certificate),
$x$ {\it forgets} $a$ at $t'$ such that $t \lessapprox t'$
if $a$ is considered invalid, i.e., if one of the following reasons apply:
\begin{enumerate}
\item $k$ cannot be obtained ($k$ is not certified by a certificate).
\item The certificate for $k$ is invalid on $a$,
i.e., the certificate is valid since $t_{ini}$ and until $t_{exp}$, and
there exists time $t'' \sqsubset a$ such that $t'' < t_{ini}$ or
$t_{exp} < t''$.
\item $sig(a_2, k^{-1})$ is not verified with $k$.
\end{enumerate}
Moreover, if $a$ is invalid and $a \sqsubset a'$, then $a'$ is also invalid
(applicable to the case of {\em blocks} to be described in
section~\ref{subsubsec-blockchain}).

If none of the reasons above applies, then $x$ {\it validates} $a$.

\subsubsection{Services and Clients}
A service $s \in E_{serv}$ provides a readout function for
entity $e \in E$, which takes $H(n, c.id)$ as its input,
and returns, when called at $t$,
$\{ro \:|\: ro \in RO^s_t \land c.id \sqsubset ro\}$,
where $n$ is shared between $e$, $s$ and some $r$ such that $OwnerOf(r) = s$,
and $c$ is a tag known to $e$.
This function is provided for all $e \in E$ through remote procedure
call\footnote{
This is just for keeping our arguments simple, and there can be many search
options in reality.
}:
$e \xrightarrow[t]{H(n, c.id)} s \implies
e \xleftarrow[t']{\{ro \:|\: ro \in RO^s_t \land c.id \sqsubset ro\}} s$
where $t \lessapprox t'$.

\subsubsection{Blockchain}\label{subsubsec-blockchain}
All $b \in B$ collectively maintain $EV_t$ and $CERT_t$ for all $t \in T$,
while satisfying all three blockchain properties below.
It is assumed that the majority\footnote{Some might argue that it should not
be a majority on numbers, but a majority on computational power, but it is a
question of answering the queries correctly, which is independent from block
creation.} of $b$ is correct.

\paragraph{BP-1
(only an authorized user can cause a state transition that is
allowed in the state machine)}
If $b$ {\it knows} term $a \in EV \cup CERT$ at time $t$ such that $a$ is
{\it validated} (the original sender of $a$ is authenticated by a digital
signature), then $b$ lets all $b' \in B$ such that $b' \neq b$ {\it know} $a$
eventually by transitively communicating $b \xrightarrow[t']{a} b''$ for all
direct neighbor $b''$ of $b$ such that $\neg$($b''$ {\it knows} $a$),
where $t \lessapprox t'$.

\paragraph{BP-2
(such a state transition always occurs if the authorized user
wants it to happen)}
Eventually, all correct $b \in B$ {\it knows} $EV_{t'}$ and $CERT_{t'}$ at
time $t$ where $t' \lnsim t$, so that all correct $b$ works symmetrically upon
$EV_{t'}$ and $CERT_{t'}$ while others may have stopped, continuing the
evidence services.

\paragraph{BP-3
(once a state transition occurs, it is virtually irreversible)}
For $t'$ and the smallest $t$ such that $t' < t$ and
$(EV_t \backslash EV_{t'})\cup(CERT_t \backslash CERT_{t'}) \neq \phi$,
there is a (non-empty) block $bk$ created at time $t$, denoted as $bk_t$,
which contains the set differences.
All such $bk$ (including empty ones) is in $A$, and $bk$ is communicated
across all $b \in B$ just as an evidence and a certificate are.
If part of $bk$ is invalid, then $bk$ is invalid and {\it forgotten}
(discarded) in the process of dissemination.

For any $bk_{t_1}$ and $bk_{t_2}$ such that $t_1 < t_2$, if $bk_{t_3}$ does not
exist such that $t_1 < t_3 < t_2$, then $H(bk_{t_1}) \sqsubset bk_{t_2}$, and
each $bk$ costs to be created or modified (e.g., by imposed proof of work).
The accumulated cost for modifying this structure makes $bk_t$ protected from
modification at $t'$ such that $t \lnsim t'$, as illustrated in
Fig.~\ref{fig-blockchain}.

\paragraph{Block creation time}
There is a function that returns the smallest time $t$ for term
$a$ such that $a \in EV_t \cup CERT_t$ (i.e., block creation time for $a$),
$BCT : EV \cup CERT \rightarrow T_{bc}$ where $T_{bc} \subset T$.
$T_{bc}$ is the set of block creation times.
$EV_t$ and $CERT_t$ are defined as $\{ev \in EV \:|\: BCT(ev) \leq t\}$ and
$\{cert \in CERT \:|\: BCT(cert) \leq t\}$ using the function, respectively.

\paragraph{Proof of existence}
A block $bk$ is associated with a Merkle tree built from the set
$\{ev \in EV \:|\: ev \sqsubset bk\} \cup
\{cert \in CERT \:|\: cert \sqsubset bk\}$,
where the structure of such a tree is uniquely determined by the
order of $ev$'s and $cert$'s appearing in $bk$.
$bk$ contains the Merkle root of the tree, denoted as $bk.root$.
The Merkle proof for existence of term $a \in EV \cup CERT$ contained in $bk$
is denoted as $bk.proof(a)$.
If $a \not\sqsubset bk$, then $bk.proof(a) = \phi$.

There is a function that takes block creation time $bct \in T_{bc}$ as input,
and returns the Merkle root $bk_{bct}.root$.
There also is a function that takes $\langle a, bct \rangle$ as input, where
$a \in EV \cup CERT$, and returns the Merkle proof $bk_{bct}.proof(a)$.
These functions are provided for all $e \in E$ through remote procedure calls:
$e \xrightarrow[t]{bct} b \implies e \xleftarrow[t']{bk_{bct}.root} b$ and
$e \xrightarrow[t]{\langle a, bct \rangle} b \implies
e \xleftarrow[t']{bk_{bct}.proof(a)} b$, respectively,
where $t \lessapprox t'$.
While there assumed to be some faulty or malicious $b \in B$, $e$ can be
certain of the proof of existence by asking several different $b$'s, and/or
being part of the blockchain network and having part of the hash chain
structure (which requires communication with several different $b$'s).

\paragraph{Evidence services}
Upon using the following evidence services, $e \in E$ can test authenticity of
the results returned from $b \in B$ by utilizing the proof of existence above.

A blockchain node provides an evidence function that takes $H(n, c.id)$ as its
input, and returns, when called at $t$,
$\{\langle ev, bct \rangle \:|\:
ev \in EV_t \land H(n, c.id) \sqsubset ev \land bct = BCT(ev)\}$,
where $n$ is not shared with $b \in B$, and $c$ is a tag known to $e$.
This function is provided for all $e \in E$ through remote procedure
call\footnote{
These are just for keeping our arguments simple, and there can be many search
options in reality.
}:
$e \xrightarrow[t]{H(n, c.id)} b \implies
e \xleftarrow[t']{\{\langle ev, bct \rangle \:|\:
ev \in EV_t \land H(n, c.id) \sqsubset ev \land bct = BCT(ev)\}}~b$,
where $bct < t$ for all possible $bct$ and $t \lessapprox t'$.

A blockchain node also provides a certification function that takes $H(k_r)$
as its input, and returns $\langle cert, bct\rangle$, where $cert \in CERT_t$,
$k_r \sqsubset cert$ and $bct = BCT(cert)$.
This function is provided for all $e \in E$ through remote procedure
call\footnotemark[5]:
$e \xrightarrow[t]{H(k_r)} b \implies
e \xleftarrow[t']{\langle cert, bct\rangle} b$
where $t \lessapprox t'$.

\subsection{Verification of the Model}

We give casual proofs that all propositions we presented in 
section~\ref{sec-problem} are true with the proposed model.
We do it in the reverse order, as the propositions with larger numbers are
the basis for the ones with smaller ones.
As an additional condition, for some past time $t'$,
the model requires that $t' \lnsim t$ because it relies on blockchain.

\subsubsection{Proposition 6 : Alibi proof}
{\em Non-existent data in the past cannot be fabricated as if it existed, even
after related private keys or the digital signature algorithm are compromised,}
i.e.,
if a malicious participant $m$ discovers $k^{-1}$, and make term $a$ at $t$
such that
\begin{enumerate}
\item $a$ is a readout, $k = k_r$ for $r \in R$, and $t' \sqsubset a$ such
that $t' \lnsim t$, and there exists a certificate $cert$ for $k_r$ such that
$\langle t_{ini}, t_{exp}\rangle \sqsubset cert$ and
$t_{ini} \lnsim t' \lnsim t \lnsim t_{exp}$,

\item $a$ is an evidence for such a readout above, or

\item $a$ is a certificate, $k = k_v$ for $v \in V$, and
$\langle t', t_{exp}\rangle \sqsubset a$ such that
$t' \lnsim t \lnsim t_{exp}$, and there exists a PKI certificate for making
$k_v$ valid at $t'$ and $t$,
\end{enumerate}
then $e \in E$ does not believe that $a$ existed.

\paragraph{Proof sketch}
We prove that the proposition is true by case study:
\begin{enumerate}
\item In case $a$ is a readout, because of atomicity of readout and evidence,
there needs to be the corresponding evidence.
If there is not, $a$ did not exist.
If there is, continued to the next case.

\item In case $a$ is an evidence or a certificate, because $t' \lnsim t$,
$BCT(a)$ needs to return $t''$ such that $t' \lessapprox t'' \lnsim t$.
However, because of BP-3, $bk_{t''}$ is protected from modification at around
$t$.
Thus $BCT(a)$ can only return $t''$ such that $t \approx t''$ instead,
or otherwise proof of existence fails.
Therefore $a$ did not exist at $t'$.
\end{enumerate}

\subsubsection{Proposition 5 : Elapsed-time proof}
{\em Past data remains authentic even after related private keys or the digital
signature algorithm are compromised, or public key certificates are revoked or
expired.
Furthermore, any attempt to delete past authentic data will fail,}
i.e., if a malicious or faulty participant $m$ let all other participants
{\it know} $k^{-1}$ at around time $t$, and there is a past term $a$ such that
\begin{enumerate}
\item $a$ is a readout, $k = k_r$ for $r \in R$, and $t' \sqsubset a$ such
that $t' \lnsim t$, and there exists a certificate $cert$ for $k_r$ such that
$\langle t_{ini}, t_{exp}\rangle \sqsubset cert$ and
$t_{ini} \lnsim t' \lnsim t_{exp} \lnsim t$,

\item $a$ is an evidence for such a readout above, or

\item $a$ is a certificate, $k = k_v$ for $v \in V$, and
$\langle t', t_{exp}\rangle \sqsubset a$ such that
$t_{exp} < t$, and there exists a PKI certificate for making $k_v$ valid
at $t'$ but not any longer at $t$,
\end{enumerate}
or if there is an attempt to delete such $a$,
then $e \in E$ still believes that $a$ existed and was valid at $t'$.

\paragraph{Proof sketch}
Because Proposition 6 is true, no one can have a term that pretends to
have existed at around time $t'$ fabricated at $t$ such that $t' \lnsim t$.
Therefore it is inferred that $a$ existed at $t'$.
Moreover, validity of $a$ at $t'$ is verified as follows:
\begin{enumerate}
\item In case $a$ is a readout at $t'$, it is inferred that the data was
correctly signed at $t'$ at which time $k_r$ was valid.
But further verification by checking the corresponding evidence is required,
and therefore continued to the next case.

\item In case $a$ is an evidence that existed at around $t'$, with successful
proof of existence that shows $a \sqsubset bk_{t''}$ where $BCT(a) = t''$
and $t' \lessapprox t''$,
its signature is verified as above case (it is the same signature
over the same data).
Therefore $a$ was valid at $t'$.

\item In case $a$ is a certificate that existed at around $t'$, with successful
proof of existence that shows $a \sqsubset bk_{t''}$ where $BCT(a) = t''$
and $t' \lessapprox t''$, it is inferred that
$\langle k_r, t', t_{exp}\rangle \sqsubset a$ was correctly signed at or
before $t'$ at which time $k_v$ was valid.
Therefore $a$ was valid at $t'$.
\end{enumerate}
Furthermore, because of BP-3, $bk_{t''}$ is protected from modification at
around $t$, and an attempt to delete $a$ from $bk_{t''}$ will fail.

\subsubsection{Proposition 4 : Evidence service integrity}
{\em If the additional entity (blockchain) lies about the content or existence
of a digital evidence, it is detected by the client,}
i.e., all of the following
statements are true when $e \in E$ requests blockchain at time $t$ for
evidences or a certificate with $H(n, c.id)$ or $H(k_r)$, respectively:
\begin{enumerate}
\item Case of an existing and correct term :
if there exists $ev \in EV_t$ such that $H(n, c.id) \sqsubset ev$ or
$cert \in CERT_t$ such that $H(k_r) \sqsubset cert$, and if incorrect $b \in B$
does not return $ev$ or $cert$, then $e$ finds out.

\item Case of a non-existent or tampered term :
if incorrect $b$ returns $ev \not\in EV_t$ or $cert \not\in CERT_t$, then
$e$ finds out.

\item Case of an existing but wrong term :
if there exists $ev \in EV_t$ such that $H(n, c'.id) \sqsubset ev$ where
$c' \neq c$ or $cert \in CERT_t$ such that $H(k_{r'}) \sqsubset cert$ where
$r' \neq r$, and if incorrect $b \in B$ returns $ev$ or $cert$, then $e$ finds
out.
\end{enumerate}

\paragraph{Proof sketch} 
Because majority of $b \in B$ is correct and Proposition 5 is true,
$e$ will receive $ev$ or $cert$ from correct $b$'s with successful proof of
existence in the case of an existing and correct term, and therefore $e$ finds
out about incorrect $b$'s.

Because Proposition 6 is true, proof of existence fails
in the case of a non-existent or tampered term, and therefore $e$ finds out.

Because $e$ {\it knows} $H(n, c.id)$ or $H(k_r)$ as the search key, $e$ will
discover that $H(n, c.id) \not\sqsubset ev$ or $H(k_r) \not\sqsubset cert$
in the case of an existing but wrong term.

\subsubsection{Proposition 3 : Service confidentiality}
{\em No secret between the service and the client is shared with the
additional entity that provides digital evidences (blockchain),} i.e.,
for all $a \sqsubset ev$ for all $ev$ such that $r \xrightarrow[t_{ev}]{ev} b$,
$a$ is not a secret between client $e$ and service $s$, not to be shared with
$b$ (it is self-evident that no secret is shared in the cases of a certificate,
and input arguments for remote procedure calls involving an evidence are
among possible $a$'s).

\paragraph{Proof sketch}
$ev$ is in the form
$\langle h_1, h_2, sig(\langle h_1, h_2 \rangle, k^{-1}_r),
H(k_r) \rangle$, where $h_1 = H(n, c.id)$, and
$h_2 = H(t, r.loc(), c.data(t))$,
while secrets between $e$ and $s$ are
$n$, $c.id$, $t$, $r.loc()$ and $c.data(t)$.
Because $H$ is a one way function, $ev$ contains no such secret.

\paragraph{Discussion}
Although because $t \lessapprox t_{ev}$, $t$ can be approximated.
Also, $r.loc()$ may be inferred from some external clues related with $k_r$,
in which case we may have to worry about preimage attacks on $c.data(t)$.
However, we might as well worry about collusion between $e$ or $s$ and $b$,
which should be easier.
Traffic analyses by using appearance patterns of $H(k_r)$ or $r$'s IP
addresses may also have to be worried about.

\subsubsection{Proposition 2 : Service integrity}
{\em If the service lies about the content or existence of data, it is
detected by the client,}
i.e., all of the following statements are true when $e \in E$ requests
$s \in E_{serv}$ at time $t$ for readouts with $H(n, c.id)$:
\begin{enumerate}
\item Case of an existing and correct readout :
if there exists $ro \in RO^s_t$ such that $c.id \sqsubset ro$, and if incorrect
$s$ does not return $ro$, then $e$ finds out.

\item Case of a non-existent or tampered readout :
if incorrect $s$ returns $ro \not\in RO^s_t$, then $e$ finds out.

\item Case of an existing but wrong readout :
if there exists $ro \in RO^s_t$ such that $H(n, c'.id) \sqsubset ro$ where
$c' \neq c$, and if incorrect $s$ returns $ro$, then $e$ finds out.
\end{enumerate}

\paragraph{Proof sketch}
We prove that the proposition is true by case study:
\begin{enumerate}
\item Case of an existing and correct readout :
because Proposition 4 is true, for every correct $ro$, there exists the
corresponding evidence $ev$.
$e$ finds out that $s$ is incorrect if there exists $ev$ that does not match
readouts $e$ {\it knows}.

\item Case of a non-existent or tampered readout :
this is further divided into two categories as below.
\begin{enumerate}
\item Non-existent or invalid readout : there is no corresponding $ev$, so that
$e$ finds out.
\item Valid tampered readout :
this represents the cases of false time $t \sqsubset ro$ and false location
$r.loc() \sqsubset ro$ returned by $r$ without tampering $r$, somehow fooling
the positioning system (therefore small possibility).
If $t \nsim t'$ for real time $t'$, then $e$ finds out by $BCT(ev)$ for the
corresponding $ev$.
False $r.loc()$ is more difficult to detect, but we have clues :
statistical analysis using other readouts from $r$ (if only part of readouts
from $r$ have false locations), or network routes of evidences from $r$ to
$b \in B$ (which need to be passed to $e$ somehow).
\end{enumerate}

\item Case of an existing but wrong readout :
because $e$ {\it knows} $H(n, c.id)$ as the search key, $e$ will discover that
$H(n, c.id) \not\sqsubset ro$.
\end{enumerate}

\subsubsection{Proposition 1 : Intra-service confidentiality}
{\em No secret between the service and the reader is shared with the client,
yet the client can verify the authenticity of data from the reader,} i.e.,
for all $a \sqsubset ro$ for all $ro$ such that $ro$ is a readout from reader
$r$ that $s$ send to $e$, $a$ is not a secret between $s$ and $r$, not to be
shared with $e$.

\paragraph{Proof sketch}
$ro$ is in the form 
$\langle n, c.id, t, r.loc(), c.data(t),
sig(\langle h_1, h_2\rangle, k_r^{-1}), H(k_r) \rangle$.
No subterm of $ro$ is a secret between $s$ and $r$,
and still, $e$ can determine the validity of $ro$ by checking the certificate
for $k_r$, calculating $\langle h_1, h_2\rangle$, and verifying
$sig(\langle h_1, h_2\rangle, k_r^{-1})$.

\section{Feasibility}\label{sec-feasibility}
\subsection{Outline of the Prototype}
\begin{figure}[h]
\begin{center}
\includegraphics[scale=0.45]{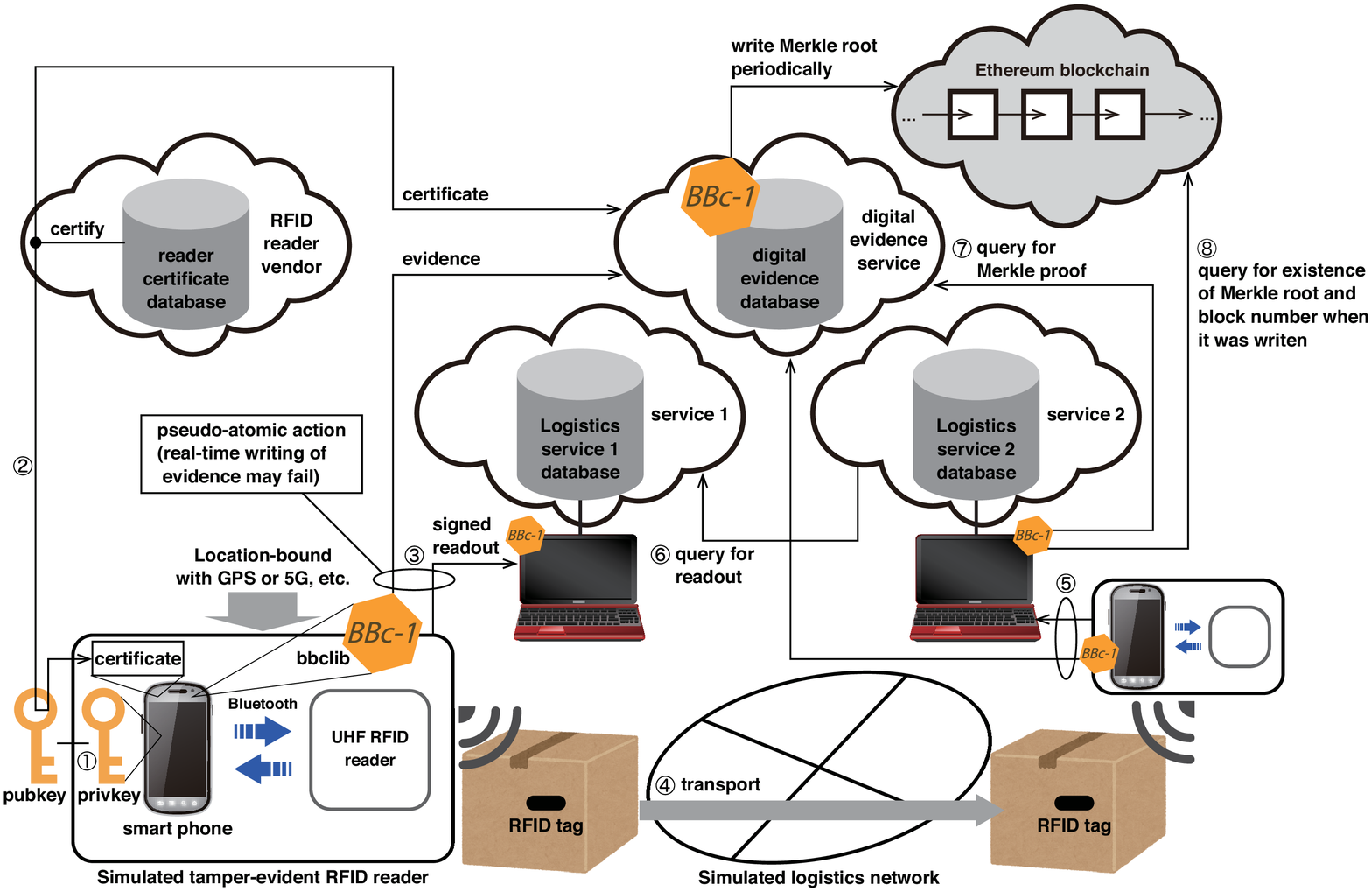}\\
{\footnotesize
\begin{itemize}
\item[*] pubkey $=$ public key, privkey $=$ private key,
bbclib $=$ library to use BBc-1 functionality
\item[*] Smart phones are chosen for ease of handling.
They can be replaced by microcomputers.
\end{itemize}
}
\caption{Experimental Prototype Being Built for Evaluation}\label{fig-exp}
\end{center}
\end{figure}

We are building an experimental prototype as illustrated in Fig.~\ref{fig-exp}
to implement our design for evaluation and demonstration purposes.
A bluetooth RFID reader is coupled with a smart phone (can be replaced by a
microcomputer) to simulate a more intelligent reader.
In the real setting, the reader is to be equipped with physical identification
as described in \cite{Watanabe2017:FTC} and \cite{Watanabe2019:ChipLevel} for
tamper-evidentness.
BBc-1\cite{Saito2017:BBc} is a toolkit to make applications of blockchain.
As can be inferred from the description in section~\ref{sec-gas-cost}, if the
evidence is written to Ethereum each time when a large number of tags are read
or written, the cost will be enormous.
Therefore, we construct a Merkle tree {\em off-chain} using this toolkit,
and periodically write only the Merkle root to Ethereum to realize a
blockchain service as a whole.

An example run of the prototype is as follows:
\begin{enumerate}
\item An RFID reader generates a key pair.

\item The vendor of the reader issues a certificate for the
relevant public key.
\begin{itemize}
\item Typically this is followed by shipment of the reader.
\end{itemize}

\item Reader reads the tag's ID and writes data (that can identify service 1
and the final destination, for example) to the tag by the order from service 1.
\begin{itemize}
\item At the same time, the corresponding evidence is sent to the digital
evidence service by the reader, and a Merkle tree is grown, and the Merkle
root is periodically written to Ethereum.
\end{itemize}

\item The package with the tag is transported to a distant relay point.

\item The tag in question is read by another reader at the relay point by the
order from service 2, yielding ID and data.
\begin{itemize}
\item At the same time, the corresponding evidence is sent to the digital
evidence service by the reader, and eventually stored in Ethereum in the form
of a Merkle root.
\end{itemize}

\item Service 2 identifies service 1 from the data, and gets the past readout
from service 1.

\item Service 2 gets the Merkle proof of the evidence regarding the past
readout from the digital evidence service.

\item Service 2 recalculates the Merkle tree from it and finds that the
readout obtained from service 1 is authentic as it matches the Merkle root
stored in an Ethereum block, and from the block number representing the block
creation time for the Merkle root, service 2 can confirm approximately by what
time the readout was recorded.
Then service 2 can forward the package to the verified final destination.
\end{enumerate}

\subsection{Pseudo-Atomicity and Manufacturing Cost of Readers}
Since this is a distributed system, it is difficult to say that transmissions
will always succeed, as failures are inevitable.
Therefore, in order to achieve pseudo-atomicity, a reader must have the
ability to store readouts and evidences in memory and retransmit them.
Such memory should be non-volatile.
Necessities of this and tamper evidentness increase the cost of
manufacturing the reader.

However, these costs should be justified because if a reader is not protected,
the reading and writing of thousands or millions of tags will not be secure.

\subsection{Window Before Digital Evidence is Verifiable}
Since proof of existence is not established until a Merkle root is written
into Ethereum, the integrity of the digital evidence service that creates the
Merkle trees is a challenge.
It is necessary to ensure that no fraud occurs during the creation of a
Merkle tree for, say, 1$\sim$2 hours before the Merkle root is written
(justification for this time window is explained in the next subsection).

One idea is for the digital evidence service to issue a short-term, bulk proof
with its own digital signature for a specified set of evidences, which
certifies that the evidences are included in the Merkle tree being made.
The logistics service obtains such a short-term proof from the digital evidence
service that there is an evidence corresponding to a group of readouts read 
in the past few minutes, for example, by the readers owned by the service.
The short-term proof is reconfirmed after the Merkle root is written, by
obtaining a proper proof of existence from the digital evidence service and
Ethereum.

\subsection{Cost of Gas}\label{sec-gas-cost}
BBc-1 provides an {\em anchoring} smart contract as shown in
Fig.~\ref{fig-bbcanchor} that runs on both Ropsten test
network and main network of Ethereum to store specified digests in blockchain.
Deploying the contract costs 149,119 gas, which is one time cost.
{\em BBcAnchor.store()} that stores a digest takes a near constant cost
of 44,241 gas on average. No gas is required for {\em BBcAnchor.isStored()} or
{\em BBcAnchor.getStored()}.

\begin{figure}[h]
\begin{center}
{\tiny
\begin{lstlisting}[frame=single]
contract BBcAnchor {
  mapping (uint256 => uint) public _digests;
  constructor () public {
  }
  function getStored(uint256 digest) public view returns (uint block_no) {
    return (_digests[digest]);
  }
  function isStored(uint256 digest) public view returns (bool isStored) {
    return (_digests[digest] > 0);
  }
  function store(uint256 digest) public returns (bool isAlreadyStored) {
    bool isRes = _digests[digest] > 0;
    if (!isRes) {
      _digests[digest] = block.number;
    }
    return (isRes);
  }
}
\end{lstlisting}
}
{\footnotesize
\begin{itemize}
\item[*] This saves the current block number for a stored digest.
\end{itemize}
}
\caption{BBcAnchor Smart Contract Code (excerpt)}\label{fig-bbcanchor}
\end{center}
\end{figure}

Table~\ref{tab-confirm} shows recommended gas prices from \cite{COC:EGS}
in Gwei ($10^9$ wei, where 1 ETH $= 10^{18}$ wei) and estimated mean time to
confirm in the case 44,241 gas were used in a transaction.
The table does not list prices in a fiat currency because they vary.

\begin{table}[h]
\begin{center}
\caption{Gas Price and Estimated Mean Time to Confirm One Write}\label{tab-confirm}
\begin{tabular}{l|r|r}\hline
\multicolumn{1}{c|}{Policy}&
\multicolumn{1}{c|}{Gas Price (Gwei)}&
\multicolumn{1}{c}{Mean Time to Confirm (sec)}\\\hline
Fastest&	85&			26$\sim$27\\\hline
Average&	50$\sim$53&	269$\sim$299\\\hline
Cheap&		33&			1,091$\sim$1,140\\\hline
\end{tabular}
\end{center}
\end{table}

\begin{figure}[h]
\begin{center}
\includegraphics[scale=0.4]{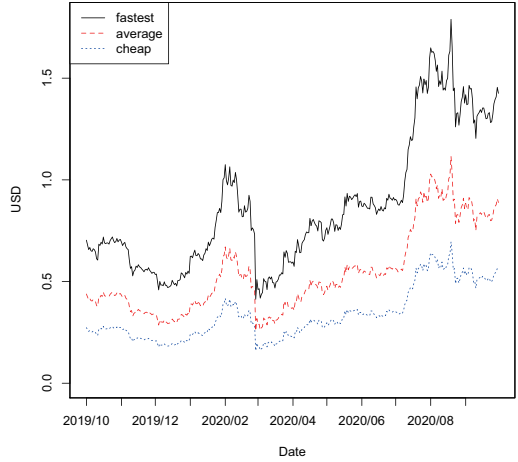}\\
\caption{Estimated Cost of Gas per Storing a Digest for the Past One Year}
\label{fig-gas-cost}
\end{center}
\end{figure}

Now we calculate the cost in USD using market prices of ETH for the past one
year obtained from \cite{Etherscan:Charts}.
As we see no apparent correlation between ETH price and average gas price
in ETH from the charts provided in \cite{Etherscan:Charts}, we just stuck to
the recommended prices of gas in Gwei as of October 2020 throughout the past
one year to obtain the rough estimate in Fig.~\ref{fig-gas-cost}.

Assuming a domestic logistics service consisting of roughly a single time zone,
and an average of one hour of travel time between locations during the day and
two hours at night, we propose 18 writes of Merkle roots per day for seven days
a week for estimation.
Table~\ref{tab-cost} shows the estimated annual cost of gas under the
condition.

\begin{table}[h]
\begin{center}
\caption{Estimated Total Annual Cost of Gas in USD with 18 Writes a Day}\label{tab-cost}
\begin{tabular}{l|r}\hline
\multicolumn{1}{c|}{Policy}&
\multicolumn{1}{c}{USD/year}\\\hline
Fastest&	5,705\\\hline
Average&	3,557\\\hline
Cheap&		2,215\\\hline
\end{tabular}
\end{center}
\end{table}

\begin{figure}[h]
\begin{center}
\includegraphics[scale=0.4]{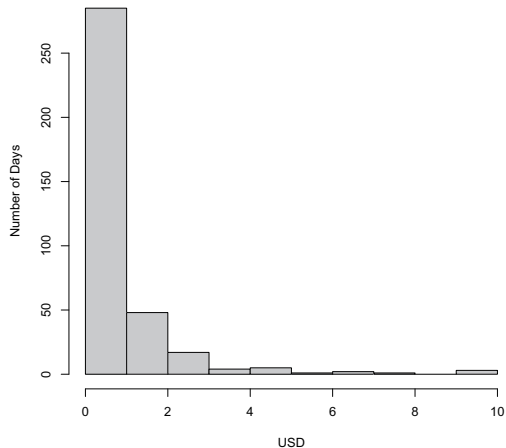}\\
\caption{Histogram of Cost per One Write with Actual Average Gas Prices for the Past One Year}
\label{fig-average-gas-cost}
\end{center}
\end{figure}

Readers may wonder that more accurate estimation of the cost could be done
with average daily gas prices available from \cite{Etherscan:Charts}.
But their chart suggests occasional surges that probably should be considered
outliers.
Fig.~\ref{fig-average-gas-cost} shows frequencies of cost for storing a digest
based on the daily average gas prices and ETH prices for the past one year.
We see that the cost is mostly below 1 USD.
We get total annual gas cost of 4,595 USD using this estimation, which should
be higher than the practical average, although as a business entity, the
digital evidence service would more incline for the {\em fastest} policy
to pay higher prices.

In any case, we estimate that the annual cost of gas roughly ranges between
2,000$\sim$6,000 USD.
This should be insignificant in comparison to the whole cost for operating
a nation-wide digital evidence service.
The results show that genuine logistical information can be shared with
moderate cost.
The cost would be more than offset by the increased benefits this
infrastructure will bring.

\section{Discussion}\label{sec-discussion}
\subsection{More Applications}
Let us consider some socially impactful applications of our proposal here,
in addition to extensive and genuine logistics monitoring.
There are some research on monitoring temperature\cite{Jedermann2009145}
or vibration\cite{Yang8125166}\cite{Li8946594} of packages being shipped
by constant monitoring with RFID readers.
This kind of monitoring is necessary and is allowed to be reasonably costly
when transporting expensive wine, art, virus specimen, precision equipment,
etc.

But a capacitive sensor can be configured to leave a record in the memory of
a passive tag when a certain physical quantity is exceeded.
Then, with our proposal, we may be able to obtain digital evidences of
packages (not) exceeding certain level of temperature or vibration during
shipment just by reading the tag at the destination without constantly
monitoring these physical parameters during shipment, providing an inexpensive
and thus more widely applicable alternative for such monitoring.

\subsection{Generalization}
\begin{figure}[h]
\begin{center}
\includegraphics[scale=0.6]{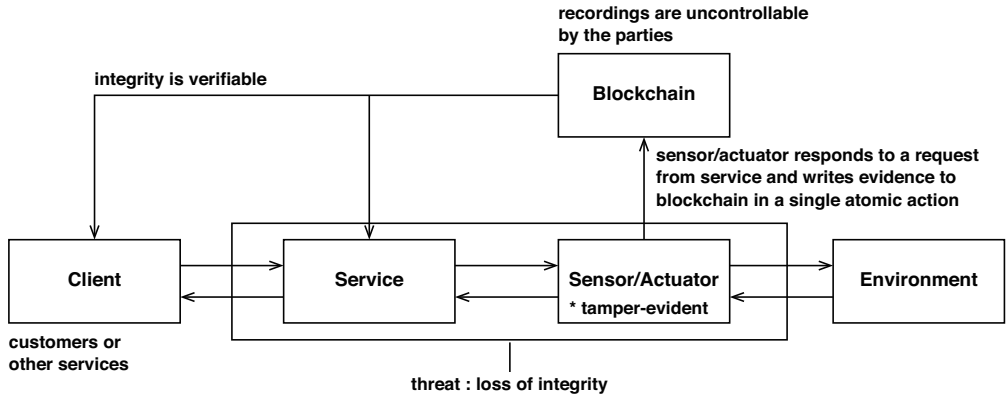}\\
\caption{Overview of the Design for General IoT Applications}
\label{fig-design-iot}
\end{center}
\end{figure}

Our model can be extended to fit general IoT (Internet of Things) as shown in
Fig.~\ref{fig-design-iot}, by abstracting the reader/writer as a
sensor/actuator.
Although the above mentioned temperature and vibration sensing is one variant
of this model (i.e., a sensor is placed in the environment in the form of a
tag), we think that this model can also be applied to other types of sensors
or actuators in more straightforward ways, including surveillance cameras,
activity logs, robots, environmental controllers, etc.

In addition to increasing the trust of individual applications, we believe
that this general-purpose infrastructure will allow genuine data to be shared
and aggregated across services, making it possible to create appropriate
datasets to train neural networks, for example, and contribute to optimizing
the whole system.

\section{Related Work}\label{sec-related-work}
\subsection{Digital Evidence with Blockchain}
Block-DEF\cite{Tian2019151} is a framework for blockchain-based digital
evidence services.
It shares a commonality with our proposal to store large amount of evidences
in blockchain, and make them verifiable.
However, whereas Block-DEF is trying to create a new blockchain, our proposal
can use an existing blockchain to meet scalability and other properties.
In particular, state machine replication based on PBFT (Practical Byzantine
Fault Tolerance) by a private group of participants, the method used in
Block-DEF, requires an estimate of the at most number of incorrect nodes, as
pointed out by \cite{Saito2016:Blockchain}, which may assume probabilistic
behaviors of participants.
It is unlikely to be suitable for evidence services because deliberate attacks,
collusion, or takeover is not a question of probability but intention.
This means, as \cite{Saito2020:Blockchain} points out, that there is no way to
verify the authenticity of a replica that each participant has because they
may be fooled by others, which is more difficult to do with a public platform
like Ethereum.

\subsection{RFID with Blockchain}
Authentication among RFID tags, readers and services where blockchain is used
in a supply chain network is studied in \cite{Sidorov8598865}, which might
complement our proposal.
However, likewise, their work assumes private group of participants for
blockchain where those participants need permissions to join, so that it must
face the difficulty of proving to the outside world the correctness of the
replicas maintained in a controlled membership without trusting the members.

\subsection{(Post) Supply Chain Management with Blockchain}
Post supply chain management using Ethereum is studied in \cite{Toyoda7961146},
which is an interesting problem to think about because we can conceive that
products can be RFID-tagged until they reach the hands of the consumer.
We think that our proposal fits for this situation as well, because it can be
applied where there is a reader, and we think that it is more scalable because
it keeps the Merkle tree of records off-chain.

A token recipes model\cite{Westerkamp8726739} has been proposed that uses
digital tokens on blockchain for the purpose of tracking the products created
and processed from the ingredients.
This raises an important issue, but it is under the influence of the oracle
problem (see section~\ref{subsec-blockchain}), and probably won't be solved
unless it is considered along with manufacturing automation.

Applications of blockchain to the food supply chain is of high interest as it
relates to our healthy lives.
One trial is \cite{Mondal8674550}, but again, the records seem to be maintained
by a private group of agents, facing the problem of trust.
\cite{Tian7538424} and \cite{Tian7996119} give only high-level descriptions
of food supply chain applications, but they show specific problems to think
about when applying the digital evidence technology to food supply chain,
such as the list of safety risks with respect to foods.

\section{Conclusions}\label{sec-conclusions}
Although we will continue to work on the prototype for evaluation, and
refine our proposal, our contributions so far are as follows:
\begin{enumerate}
\item We proposed a design of an information infrastructure in which the
information on tags read and written at controlled locations by controlled
RFID readers are correctly shared in the logistics network.
\item We made a series of propositions that need to be true in order for such
a design to work as intended.
We made a semi-formal model of our proposal, and verified that all the
propositions are true under the model.
However, we may have to be alerted for the possible traffic analyses by the
digital evidence service in order to keep confidentiality of the logistics
services.
\item We introduced a prototype design, and evaluated feasibility of our
proposal.
In particular, we estimated the cost of operating a smart contract on Ethereum,
and found that that would impose just a moderate cost onto the digital
evidence service, which would be more than offset by the increased benefits
this infrastructure will bring.
\end{enumerate}

This makes it possible to trace authentic logistics information using
inexpensive passive RFID tags.
Furthermore, by abstracting the reader/writer as a sensor/actuator, the method
can be extended to the Internet of Things (IoT) in general.

\bibliographystyle{plain}
\bibliography{logistics-blockchain}

\end{document}